\documentclass[11pt]{article}
\usepackage{amsmath}
\usepackage{amssymb}
\usepackage{amsthm,amsxtra}
\usepackage{epsf}
\usepackage{eepic}
\newlength{\mytopmargin}
\newlength{\myleftmargin}
\newtheorem{prop}{Proposition}
\begin{document}
\vspace{4cm}
\noindent
{\bf \Large Derivation of an eigenvalue probability density
function relating to the Poincar\'e disk} 

\vspace{5mm}
\noindent
Peter J.~Forrester${}^*$
 and Manjunath Krishnapur${}^\dagger$

\noindent
${}^*$Department of Mathematics and Statistics,
University of Melbourne, 
Victoria 3010, Australia 
${}^\dagger$
Department of Mathematics, University of Toronto, ON M5S 2E4, Canada 

\small
\begin{quote}
A result of Zyczkowski and Sommers [J.~Phys.~A {\bf 33}, 2045--2057 (2000)] gives the
eigenvalue probability density function for the top $N \times N$ sub-block of a Haar
distributed matrix from $U(N+n)$. In the case $n \ge N$, we rederive this result, starting from knowledge of the
distribution of the sub-blocks, introducing the Schur decomposition, and
integrating over all variables except the eigenvalues. The integration is done by identifying
a recursive structure which reduces the dimension. This approach is inspired by an analogous
approach which has been recently applied to determine the eigenvalue probability density function for
random matrices $A^{-1} B$, where $A$ and $B$ are random matrices with entries standard
complex normals. We relate the eigenvalue distribution of the sub-blocks to a many
body quantum state, and to the one-component plasma, on the pseudosphere. 
\end{quote}

\section{Introduction}

The plane, sphere and pseudosphere are special geometries in a number of related many
body statistical systems. In particular the lowest Landau level wave function
for quantum
particles in a magnetic field, the equilibrium statistical mechanics of a one-component
plasma at a special value of the coupling, and the eigenvalue probability density
function for certain random matrix ensembles all give rise to a special solvable
state in these geometries. In regard to the first of these,
regarding the geometries as surfaces in appropriate 
3-spaces, and imposing a perpendicular magnetic field, the solvable state is the 
ground state wave function, in that it 
can be written in an explicit factorized (Jastrow) form and the correlations can be
expressed explicitly as determinants \cite{Co87,Du92a}. The absolute value squared of these wave
functions have the interpretation as the Boltzmann factor for a one-component plasma, confined to the
corresponding surface, and interacting at the special value of the coupling $\Gamma := q^2/k_B T = 2$
\cite{Ca81,JT98}.
And the Boltzmann factors, after projecting onto the plane via an appropriate stereographic projection
in the case of the sphere and pseudosphere, allow for realizations as the eigenvalue probability density
function for three particular random matrix ensembles \cite{Gi65,ZS99,Kr06}.
The determinant form for the correlations in the quantum problem
carries over to the plasma and eigenvalue distributions, so these many body systems are also solvable.
Although not to be addressed here, we mention too that these same geometries play a special role in
relation to the study of the zeros of random polynomials \cite{FH98,Le00}.

It is the purpose of the present paper to contribute to the theory of the random matrix ensemble relating
to the pseudosphere, or more precisely to the Poincar\'e disk. Specifically, our interest is in the
eigenvalue distribution of the ensemble of the top $N \times N$ sub-block of Haar distributed matrices
in $U(N+n)$ (due to the invariance of Haar distributed unitary matrices with respect to left and right
multiplication by permutation matrices, in fact sub-blocks formed by the entries of any $N$ rows and
$N$ columns will have the same distributions as the top $N \times N$ sub-block). This ensemble has shown
itself in a number of recent works in random matrix theory \cite{Fo06a,FK07a,Kr09,WF08}.
We present a new derivation of a
result due to Zyczkowski and Sommers \cite{ZS99}, giving that the eigenvalue probability density
function is proportional to
\begin{equation}\label{zx}
\prod_{l=1}^N (1 - |z_l|^2)^{n-1} \chi_{|z_l|<1} \prod_{1 \le j < k \le N} |z_k - z_j|^2,
\end{equation}
(here $\chi_A = 1$ if $A$ is true, and $\chi_A = 0$ otherwise). 
Our proof, although applying only to the case when 
$n\ge N$, is arguably simpler than the original. 
Moreover, our derivation unifies the case of the sphere and pseudosphere, by showing that a recently
formulated approach to the eigenvalue probability density function of the random matrix ensemble
corresponding to the sphere \cite{HKPV08}, can be adapted to the case of the pseudosphere.

\section{The pseudosphere and plasma system}
\setcounter{equation}{0}
The pseudosphere refers to the two-dimensional hyperbolic space with constant negative Gaussian
curvature $\kappa = - 1/a^2$ \cite{BV86}. It is naturally embedded in the three dimensional
Minkowski space with coordinates $(y_0,y_1,y_2)$ and line element
\begin{equation}\label{le}
(ds)^2 = - (dy_0)^2 + (dy_1)^2 + (dy_2)^2.
\end{equation}
The pseudosphere is defined as the upper branch of the equation
$$
- y_0^2 + y_1^2 + y_2^2 = - a^2
$$
and can be parameterized by
$$
y_0 = a \cosh \tau, \qquad y_1 = a \sinh \tau \cos \phi, \qquad y_2 = a \sinh \tau \sin \phi.
$$
In terms of this parametrization the line element (\ref{le}) reads
\begin{equation}\label{le1}
(ds)^2 = a^2 (d \tau)^2 + a^2 \sinh^2 \tau \, (d \phi)^2.
\end{equation}
Furthermore, the geodesic distance between $(\tau,\phi)$ and $(\tau',\phi')$ is such that
\begin{equation}\label{gd}
\cosh {s \over a} = \cosh \tau \cosh \tau' - \sinh \tau \sinh \tau'
\cos (\phi - \phi'),
\end{equation}
while the volume element corresponding to the surface area is
\begin{equation}\label{gd1}
dS = a^2 \sinh \tau \, d\tau d\phi.
\end{equation}

With $z:= x + iy$, the pseudosphere is projected onto the Poincar\'e disk via the
stereographic projection
\begin{equation}\label{gd2}
z = 2a \tanh {\tau \over 2} \, e^{i \phi}, \qquad |z| < 2a
\end{equation}
(equivalently $(y_0,y_1,y_2) \mapsto 2 a (y_1/y_0, y_2/y_0)$) and then
\begin{equation}\label{gd3}
dS = {dx dy \over (1 - |z|^2/4 a^2)^2}.
\end{equation}
Note that with $z$ expressed in polar form, (\ref{gd2}) is equivalent to writing
$$
r =  2a \tanh {\tau \over 2}, \qquad |z| < 2a.
$$

In terms of the coordinates $(\tau,\phi)$ the Laplacian $\Delta$ is computed from
(\ref{le1}) as
$$
\Delta = {1 \over a^2} \Big ( {1 \over \sinh \tau} {\partial \over \partial \tau} \sinh \tau
{\partial \over \partial \tau} + {1 \over \sinh^2 \tau}
{\partial^2 \over \partial \phi^2} \Big ).
$$
Alternatively, in terms of the coordinates $(r,\phi)$ the Laplacian reads
\begin{equation}\label{pe}
\Delta = \Big ( 1 - {r^2 \over 4 a^2} \Big )^2 \Big (
{\partial^2 \over \partial r^2} + {1 \over r} {\partial \over \partial r} +
{1 \over r^2} {\partial^2 \over \partial \phi^2} \Big ).
\end{equation}
From the Laplacian we can specify the Coulomb potential on the pseudosphere.
In fact, two potentials $\Phi$ and $\tilde{\Phi}$ will be defined. 
With $s_{jk}$ denoting the geodesic separation between particles $j$ and $k$, and
$\Delta_j$ the Laplacian with respect to the particle $j$,
the first is specified as
the solution of the Poisson equation
\begin{equation}\label{pc}
\Delta_j \Phi(s_{jk}) = - 2 \pi \delta(s_{jk})
\end{equation}
subject to the boundary condition that $\Phi(s) \to 0$ as $s \to \infty$, and the second as
the solution of the Poisson equation
\begin{equation}\label{pc1}
\Delta_j \tilde{\Phi}(s_{jk}) = - 2 \pi \delta(s_{jk}) - {1 \over 2 a^2}.
\end{equation}
In both (\ref{pc}) and (\ref{pc1}) $\delta(s_{jk})$ denotes the delta function relative to the
volume form (\ref{gd1}). The Poisson equation (\ref{pc}) corresponds to the potential at 
$(\tau_j,\phi_j)$ due to a unit charge at $(\tau_k,\phi_k)$, while
(\ref{pc1}) corresponds to the potential at 
$(\tau_j,\phi_j)$ due to a unit charge at $(\tau_k,\phi_k)$ and a uniform smeared out
charge of the same sign and density $1/4 \pi a^2$. The solution of (\ref{pc}) is readily
verified to be
\begin{equation}\label{ps}
\Phi(s_{jk}) = - \log \tanh {s_{jk} \over 2} = - \log \Big |
{(z_j - z_k)/2a \over 1 - z_j \bar{z}_k/4 a^2} \Big |,
\end{equation}
and that of (\ref{pc1}) to be
\begin{equation}\label{ps1}
\tilde{\Phi}(s_{jk}) = - \log \sinh {s_{jk} \over 2} = - \log \Big (
{|z_j - z_k|/2a \over (1 - (r_j/2a)^2)^{1/2} (1 - (r_k/2a)^2)^{1/2} } \Big ).
\end{equation}

Choosing one or other of (\ref{ps}), (\ref{ps1}), we want to form a one-component plasma system,
consisting of  $N$ mobile unit charges, and a smeared out uniform background charge density
of net charge $- \eta$. In the large $N$ limit, we know from \cite{JT98} that the bulk
equilibrium statistical mechanics is the same for both. However,
our interest is in finite $N$, and to obtain an analogy with a random matrix eigenvalue
probability density function it is necessary to base the plasma system on the potential (\ref{ps1}).
Since (\ref{ps1}) corresponds to (\ref{pc1}), to get a net background charge density $-\eta$
we must also cancel the smeared out positive charge implied by the constant term therein,
and thus impose a smeared out charge density $- \eta - N/4 \pi a^2$. This will interact with 
the particles to create a potential $V(r)$, which must satisfy the Poisson equation
$$
\Delta V(r) = 2 \pi (\eta + N/4 \pi a^2).
$$
Choosing for convenience $V(0) = 0$ (the constant terms in the total potential are of
no interest to us), and using the form (\ref{pc}) of $\Delta$, the solution of this
equation is seen to be
\begin{equation}\label{Vr}
V(r) = - (2 \pi \eta a^2 + N/2) \log \Big ( 1 - {r^2 \over 4 a^2} \Big ).
\end{equation}
The total potential energy $U$, excluding constants, is obtained by summing the potential
(\ref{ps1}) over all pairs, and adding to this 
$\sum_{j=1}^N V(r_j)$. We thus have \cite{JT98}
$$
U = - \sum_{1 \le j < k \le N} \log \Big (|z_k - z_j|/2a \Big ) -
(2 \pi \eta a^2 + 1/2) \sum_{j=1}^N \log (1 - r_j^2/4 a^2)
$$
implying that the Boltzmann factor is such that
\begin{eqnarray}\label{Vr1}
&& e^{- \beta U} dS_1 \cdots dS_N 
\propto \prod_{l=1}^N (1 - |z_l|^2/4 a^2)^{(2 \pi \eta a^2 + 1/2)\beta - 2}
\prod_{1 \le j < k \le N} |z_k - z_j|^\beta
dx_1 dy_1 \cdots dx_N dy_N.
\end{eqnarray}

\section{Magnetic analogy}
\setcounter{equation}{0}
It is remarked in \cite{JT98} that with $l := \sqrt{\hbar c/eB}$ the magnetic length
\begin{equation}\label{mn}
\psi_j(z) = \Big (1 - {|z|^2 \over 4 a^2} \Big )^{(a/l)^2} \bar{z}^j, \qquad j=0,1,\dots
\end{equation}
are, up to normalization, the lowest Landau level eigenstates for a quantum particle on the
pseudosphere, in the presence of a constant perpendicular magnetic field $B$, and projected
onto the Poincar\'e disk. Forming a Slater determinant out of the first $N$ of these
gives an $N$-body state $\psi$ such that $|\psi|^2$ is
proportional to (\ref{Vr1}) with 
\begin{equation}\label{bb}
\beta = 2, \qquad  4 \pi \eta a^2 + 1 = 2 (a/l)^2.
\end{equation}
We take this opportunity to gives some details on how (\ref{mn}) comes
about (see also \cite{Fo02}).

Let $\vec{r}$ denote the unit vector perpendicular to the pseudosphere in Minkowski space.
By the formal analogy with the surface of a sphere with radius $R$ and polar
co-ordinates $(\theta,\phi)$ as specified by
$$
a \leftrightarrow iR, \qquad \tau \leftrightarrow i \theta
$$
one sees that the vector potential
\begin{equation}\label{sp1}
\vec{A} = \Big (0,0, {Ba \over \sinh \tau} (1 - \cosh \tau) \Big )
\end{equation}
is such that $\nabla \times \vec{A} = B \hat{\vec{r}}$ and so corresponds to the sought
perpendicular magnetic field. Further, it follows similarly that the Hamiltonian
$$
H := {1 \over 2m} \Big ( - i \hbar \nabla + {e \over c} \vec{A} \Big )^2
$$
is such that
\begin{equation}\label{sp2}
H = {\hbar^2 \over 2 m a^2} \Big ( - {1 \over \sinh \tau} {\partial \over \partial \tau}
\Big ( \sinh \tau {\partial \over \partial \tau} \Big ) +
{1 \over \sinh^2 \tau} \Big (  {\partial \over \partial \phi} -
\Big ( {a \over l} \Big )^2 (\cosh \theta - 1)^2 \Big )
\end{equation}
(of course in writing down (\ref{sp1}) and (\ref{sp2}) knowledge of the corresponding
results for the sphere are assumed \cite{WY76}). The change of variables (\ref{gd2}) so as
to project to the Poincar\'e disk gives that (\ref{sp2}) reads
\begin{equation}\label{sp3}
H = {\hbar^2 \over 2m} \Big ( - 4l^2 \Big ( 1 - {|z|^2 \over 4 a^2} \Big )
{\partial^2 \over \partial z \partial \bar{z} } +
 \Big ( 1 - {|z|^2 \over 4 a^2} \Big ) \Big ( z {\partial \over \partial z} - \bar{z}
 {\partial \over \partial \bar{z} } \Big ) +
{|z|^2 \over 4 l^2} \Big ).
\end{equation}

Following \cite{Du92a}, with $\omega_c := eB/mc$, we note the projected Hamiltonian (\ref{sp3})
is of the form
\begin{equation}\label{sp4}
H = - {\hbar \omega_c \over 2} \Big ( \Big ( {\partial^2 \Phi \over \partial z  \partial \bar{z} }
\Big )^{-1} \Big ( {\partial \over \partial \bar{z} } - {\partial \Phi \over \partial \bar{z} }
\Big ) \Big ( {\partial \over \partial {z} } + {\partial \Phi \over \partial {z} } \Big ) - 1 \Big )
\end{equation}
with
$$
\Phi = - \Big ( {a \over l} \Big )^2 \log \Big ( 1 - {|z|^2 \over (2a)^2} \Big ).
$$
But (\ref{sp4}) is such that
$$
e^{\Phi} H e^{- \Phi} = - {\hbar \omega_c \over 2}
\Big ( \Big ( {\partial^2 \Phi \over  \partial {z}  \partial \bar{z}} \Big )^{-1}
\Big ( {\partial \over \partial z} - 2 {\partial \Phi \over \partial \bar{z} } \Big )
{\partial \over \partial z} - 1 \Big )
$$
and so we are lead to the result that any function of the form $e^{-\Phi} f(\bar{z})$ with $f$
analytic in $\bar{z}$ belongs to the lowest Landau level (i.e.~is an eigenfunction of $H$ with
eigenvalue $\hbar \omega_c/2$). To further specify $f$, one notes that
$$
J := z  {\partial \over \partial {z} } -  \bar{z}  {\partial  \over \partial \bar{z} }
$$
commutes with $H$ as defined by (\ref{sp4}) for general $\Phi = \Phi(|z|^2)$, and seeks
simultaneous eigenfunctions of $H$ and $J$. Such eigenfunctions are given by
$$
e^{-\Phi} \bar{z}^j, \qquad j=0,1,\dots
$$
and (\ref{mn}) follows.

Now if all the particles are fermions, the ground state $\psi$ is formed out of a Slater determinant
of the single particles states,
\begin{eqnarray}
\psi(z_1,\dots,z_N) & = & {1 \over C} \det [ \psi_{j-1}(z_j) ]_{j,k=1,\dots,N} \nonumber \\
& = & {1 \over C} \prod_{l=1}^N \Big ( 1 - {|z_l|^2 \over 4 a^2} \Big )^{(a/l)^2}
\prod_{1 \le j < k \le N} (\bar{z}_k - \bar{z}_j)
\end{eqnarray}
where $C$ is an appropriate normalization, and the second equality follows upon use of the
Vandermonde determinant identity. The modulus squared is indeed identified with the case
(\ref{bb}) of (\ref{Vr1}).

\section{Random matrix ensembles}
\setcounter{equation}{0}
\subsection{The spherical ensemble}
The eigenvalue probability density function for matrices $Y = A^{-1} B$ with $A$ and $B$ random $N \times N$ matrices consisting of independent complex Gaussian entries, turns out to be a one-component plasma on the sphere. This spherical analogue of (\ref{zx}) was proved in \cite{Kr09}.  Below, we give a simplified and unified approach that works for the sphere as well as the pseudosphere. We first sketch the proof for the sphere  (more details may be found in the forthcoming book \cite{HKPV08}). The first step is to determine the distribution of $Y$.

\begin{prop}\label{pY}
The distribution of $Y$ is proportional to
\begin{equation}\label{kay}
{1 \over \det ({\mathbb I}_N + Y Y^\dagger)^{2N} }.
\end{equation}
\end{prop}

\noindent 
Proof. \quad 
With $X$ a complex matrix and $(dX)$ denoting
the wedge product of the differentials of the real and imaginary parts of its elements,
by noting 
\begin{equation}\label{BAY}
(dB) = |\det A|^{2N} (dY)
\end{equation}
it follows that the joint density of $A$ and $Y$ is proportional to
$$
|\det A|^{2N} e^{- {\rm Tr}\{A({\mathbb I}_N + Y Y^\dagger)A^{\dagger}\}}.
$$
The distribution of $Y$ is obtained by integrating this over $A$. For this
we change variables $C = A({\mathbb I}_N + Y Y^\dagger)^{1/2}$ and by an appropriate analogue of (\ref{BAY}) and integration over $C$, obtain the density of $Y$ to be
$$
\int \frac{|\det(C)|^{2N}e^{- {\rm Tr}(CC^{\dagger})}}{\det({\mathbb I}_N+YY^{\dagger})^{N}} 
\frac{dC}{\det({\mathbb I}_N +YY^{\dagger})^{N}},
$$
from which (\ref{kay}) results. \hfill $\square$ 

\medskip
The second and final step is to use knowledge of the distribution (\ref{kay}) to proceed
to compute the eigenvalue probability density function.

\begin{prop} \label{pMM} \cite{Kr06} The eigenvalue  probability density function for $Y$ is
proportional to
\begin{equation}\label{kay1}
\prod_{l=1}^N {1 \over (1 + |z_l|^2)^{N+1} } \prod_{1 \le j < k \le N} |z_k - z_j|^2.
\end{equation}
\end{prop}

\noindent
Proof. \quad According to the Schur decomposition, $Y$ can be written
\begin{equation}\label{YU}
Y = U T_N U^{-1}
\end{equation}
where $U$ is a unitary matrix, and $T_N$ an upper triangular matrix in which the diagonal entries,
denoted $\{z_j\}_{j=1,\dots,N}$, are the eigenvalues of $Y$. Furthermore, we know that (see appendix 35 of \cite{Me91}  or section 6.3 of \cite{HKPV08})
\begin{equation}\label{YU1}
(dY) = \prod_{1 \le j < k \le N} |z_k - z_j|^2 (U^\dagger dU) \wedge_{j=1}^N dz_j
(d \tilde{T}_N)
\end{equation}
where $ \tilde{T}_N$ denotes the strictly upper portion of $T_N$. Noting from (\ref{YU}) that
$Y Y^\dagger =
U T_N T_N^\dagger U^\dagger$ shows that (\ref{kay}) is independent of $U$. The factorization of the
$U$ dependence in (\ref{YU1})
then implies that the eigenvalue distribution is
obtained from (\ref{kay}) by integrating out the dependence of $\tilde{T}_N$ and is thus
proportional to
\begin{equation}\label{YU2}
\prod_{1 \le j < k \le N} |z_k - z_j|^2 \, I_N(z_1,\dots,z_N), \qquad
I_N(z_1,\dots,z_N) := \int {1 \over \det ({\mathbb I}_N + T_N T_N^\dagger)^{2N} } \, (d \tilde{T}_N).
\end{equation}

Following \cite{HKPV08}, we will compute the integral $I_N$ by deducing a recurrence for the
integrals
\begin{equation}\label{IT}
I_{n,p}(z_1,\dots,z_n) := \int {1 \over \det ({\mathbb I}_n + T_n T_n^\dagger)^{p} } \, (d \tilde{T}_n)
\end{equation}
where $p \ge n$ (this ensures convergence). 
Analogous strategies can be found in \cite{Hu63,EKS94,FK07}.
For this purpose, let $\vec{u}$ denote the last column of
$\tilde{T}_n$. We can then write
\begin{equation}\label{TT1}
{\mathbb I}_n + T_n T_n^\dagger = \left [
\begin{array}{cc} {\mathbb I}_{n-1} + T_{n-1} T_{n-1}^\dagger + \vec{u} \vec{u}^\dagger &
z_n \vec{u} \\
\bar{z}_n \vec{u}^\dagger & 1 + |z_n|^2 \end{array} \right ],
\end{equation}
showing
$$
\det ({\mathbb I}_n + T_n T_n^\dagger) = (1 + |z_n|^2) \det \Big (  {\mathbb I}_{n-1} +
T_{n-1} T_{n-1}^\dagger + {\vec{u} \vec{u}^\dagger \over 1 + |z_n|^2} \Big ).
$$
Noting that the final term in the determinant is a matrix of rank 1, this latter formula can be
further reduced to read
\begin{equation}\label{pip}
 \det ( {\mathbb I}_n  + T_n T_n^\dagger) = (1 + |z_n|^2) \det (  {\mathbb I}_{n-1} + T_{n-1} T_{n-1}^\dagger)
\Big ( 1 + {1 \over 1 + |z_n|^2} \vec{u}^\dagger (  {\mathbb I}_{n-1} + T_{n-1} T_{n-1}^\dagger)^{-1} \vec{u}
\Big ).
\end{equation}

Hence, by first writing
$$
I_{n,p}(z_1,\dots,z_n) = \int  (d \tilde{T}_{n-1})
\int (d \vec{u} ) \, 
{1 \over \det ({\mathbb I}_n + T_n T_n^\dagger)^{p} }
$$
we can substitute (\ref{pip}) to obtain
\begin{eqnarray}\label{ICT}
&&I_{n,p}(z_1,\dots,z_n) = {1 \over (1 + |z_n|^2)^p }
\int (d \tilde{T}_{n-1}) \, {1 \over \det ({\mathbb I}_{n-1} + T_{n-1} T_{n-1}^\dagger)^p }
\nonumber \\
&& \qquad \qquad \times
\int (d \vec{u}) \,
\Big ( 1 + {1 \over 1 + |z_n|^2} \vec{u}^\dagger (  {\mathbb I}_{n-1} + T_{n-1} T_{n-1}^\dagger)^{-1} \vec{u}
\Big )^{-p}.
\end{eqnarray}
Making the change of variables
$$
\vec{v} = (1 + |z_n|^2)^{-1/2} ( {\mathbb I}_{n-1} + T_{n-1} T_{n-1}^\dagger )^{-1/2} \vec{u}
$$
we see from this that
\begin{equation}\label{ICI}
I_{n,p}(z_1,\dots,z_n) = {C_{n,p} \over (1 + |z_n|^2 )^{p-n+1} }
I_{n-1,p-1}(z_1,\dots,z_{n-1})
\end{equation}
where
\begin{equation}\label{CNP}
C_{n,p} = \int { (d \vec{v}) \over (1 + \vec{v}^\dagger \vec{v})^p }.
\end{equation}

Iterating (\ref{ICI}) with $n=2N$, $p=N$ shows
$$
I_N(z_1,\dots,z_N) \propto \prod_{l=1}^N {1 \over (1 + |z_l|^2)^{N+1} }
$$
and this substituted in (\ref{YU2}) gives (\ref{kay1}). \hfill $\square$

\medskip
We remark that the name spherical ensemble comes about because upon making the
stereographic projection from the plane to the sphere,
$$
z = 2R \tan {\theta \over 2} \, e^{i \phi}
$$
(cf.~(\ref{gd2})) transforms (\ref{kay1}) to the form
$$
\prod_{1 \le j < k \le N} |\vec{r}_j - \vec{r}_k |^2
$$
where $\vec{r}_j$ is the vector in $\mathbb R^3$ corresponding to the point
$(R,\theta_j,\phi_j)$ on the sphere.

\subsection{Truncated unitary matrices -- the pseudosphere ensemble}
Working now with truncated unitary matrices, we would like to repeat the steps which lead to
establishing the eigenvalue probability density function (\ref{kay1}). The first task then
is to specify the distribution of the top $N \times N$ sub-block of Haar distributed matrices from
$U(N+n)$. Our working requires that $n \ge N$.

\begin{prop} \cite{Co05} Let the top $N \times N$ sub-matrices in question be denoted $Q_N$. The
probability density of $Q_N$ is proportional to
\begin{equation}\label{QN}
\det ( {\mathbb I}_N - Q_N Q_N^\dagger)^{n-N},
\end{equation}
supported on $Q_N Q_N^\dagger < 1$ (i.e.~matrices $Q_N$ with all singular values less than one). 
\end{prop}

\noindent
Proof. \quad Following \cite{Fo06a}, we give a variation of the proof given in \cite{Co05}.
The proof relies on a standard result in random matrix theory giving a construction of a particular
Jacobi ensemble in terms of Wishart matrices
\cite{Mu82,Fo02}. With $c$ and $d$ being independent $N \times N$ and $n \times N$ complex Gaussian random
matrices respectively (all elements i.i.d.~standard complex normals), let $C = c^\dagger c$,
$D = d^\dagger d$. We require the fact that for $n \ge N$ the probability density of
\begin{equation}\label{J}
J := (C + D)^{-1/2} C ( C + D)^{-1/2}
\end{equation}
is proportional to
\begin{equation}\label{JJ}
\det (  {\mathbb I}_N - J )^{n-N},
\end{equation}
supported on $J < 1$.

Let $U$ denote an $(N+n) \times (N+n)$ Haar distributed unitary matrix. Let $W$ denote the $N \times (N+n)$
sub-matrix formed from the first $N$ rows, and  $\tilde{W}$ denote the $n \times (N+n)$ sub-matrix
formed from the final $n$ rows, so $U$ has the block decomposition
$$
U = \left [ \begin{array}{ll} W \\ \tilde{W} \end{array} \right ].
$$
Further, let $X$ be an $(N+n) \times N$ Gaussian matrix
of independent standard complex normals. From the facts that $WX$ is then also distributed as an
$N \times N $ Gaussian matrix of standard complex normals, and $\tilde{W}X$ as an $n \times N$ 
Gaussian matrix of standard complex normals, and that  $WX,\tilde{W}X$ are independent, we can take $c = WX$ and $d = \tilde{W}X$ in (\ref{J})
to conclude from the above result that
\begin{equation}\label{T2}
J = (X^\dagger X)^{-1/2} X^\dagger W^\dagger W X (X^\dagger X)^{-1/2}
\end{equation}
has probability density (\ref{JJ}).

Introduce the singular value decomposition by writing
$$
X = U_1 \Lambda U_2
$$
for $U_1$ an $(N+n) \times (N+n)$ unitary matrix, $U_2$ an $N \times N$ unitary matrix, and
$\Lambda$ an $(N+n) \times N$ matrix with diagonal entries (that is entries $(i,i)$, for $i\le N$) equal to the positive square root of
the eigenvalues of $X^\dagger X$. We then have
$$
X(X^\dagger X)^{-1/2} = [U_1]_{N+n,N} U_2
$$
where the notation $[Z]_{p,q}$ denotes the top $p \times q$ sub-block of $Z$. Using this in (\ref{T2}) shows
$$
J = U_2^{\dagger} [U_1^\dagger]_{N,N+n} W^\dagger W [U_1]_{N+n,N} U_2.
$$
But the distribution of $J$ is unchanged under conjugation by unitary matrices, and in particular $U_2$.
Applying this conjugation, we see that 
\begin{equation}\label{eq:equalityoflaws}
J \mbox{ has the same distribution as } [U_1^\dagger]_{N,N+n} W^\dagger W [U_1]_{N+n,N}.
\end{equation}
Since $U_{1}$ and $U$ are independent, and $U$ has Haar distribution, $UU_{1}$ has the same distribution as $U$ and hence, $W$ has the same distribution as $[WU_{1,1}\; WU_{1,2}]$, where we have written $U_{1,1}=[U]_{N+n,N}$ and $U_{1,2}$ denotes the last $n$ columns of $U$. From this, it immediately follows that $[W^{\dagger}W]_{N,N}$ has the same distribution as $[U_1^\dagger]_{N,N+n} W^\dagger W [U_1]_{N+n,N}$.  Together with \eqref{eq:equalityoflaws}, we conclude that $[W^{\dagger}W]_{N,N}$ has the same distribution as $J$.

This tells us that the probability density
of $Q_N^\dagger Q_N$ is proportional to (\ref{QN}). The probability density of $Q_N$ itself, being the top
$N \times N$ block of $U$, $U \in U(N+n)$, must be invariant under conjugation by $S \in U(N)$
and under complex conjugation, and so be
a function of (the trace of) $Q_N^\dagger Q_N$. 
But in general for a $p \times q$ ($p \ge q$) complex matrix $Y$ with the probability density $f$ of
the form $f = f(Q_N^\dagger Q_N)$, the probability density for $Y^\dagger Y$ is proportional to
$(\det Y^\dagger Y)^{p-q} f(Y^\dagger Y)$. Here $p=q=N$, so we conclude that (\ref{QN}) holds too for the probability
density of $Q_N$.
\hfill $\square$

\medskip
Knowing that $Q_N$ is distributed according to (\ref{QN}), the task now is to introduce the
Schur decomposition of $Q_N$, and integrate out all variables but the eigenvalues.
This can be done using the same recursive reduction of the dimension of the integral
as used in the proof of Proposition \ref{pMM}, and the sought probability density function
thus obtained.

\begin{prop} Let $n\ge N$. Then, the eigenvalue probability density function of $Q_N$ is proportional to (\ref{zx}).
\end{prop}

\noindent
Proof. \quad
Because $Q_N$ is a sub-block of a unitary matrix, all eigenvalues have modulus less than
one. Introducing the Schur decomposition (\ref{YU}) of $Q_N$ in (\ref{QN}), and recalling the
change of variables formula 
(\ref{YU1}), we see the eigenvalue probability density function is proportional to
\begin{equation}\label{YU2a}
\prod_{1 \le j < k \le N} |z_k - z_j|^2 \, J_N(z_1,\dots,z_N), \qquad
J_N(z_1,\dots,z_N) := \int \det ({\mathbb I}_N - T_N T_N^\dagger)^{n-N}  \, (d \tilde{T}_N),
\end{equation}
where again $\tilde{T}_N$ denotes the strictly upper triangular portion of $T_N$.

To compute $J_N$ by recurrence we introduce, for $p\ge 0$
\begin{equation}\label{pip1}
J_{m,p}(z_1,\dots,z_m) := \int \det ({\mathbb I}_m - T_m T_m^\dagger)^{p} \, (d \tilde{T}_m)
\end{equation}
(cf.~(\ref{IT})).
Analogous to (\ref{TT1}), with $\vec{u}$ denoting the last column of $\tilde{T}_N$, we can
write
$$
{\mathbb I}_m - T_m T_m^\dagger = \left [
\begin{array}{cc} {\mathbb I}_{m-1} - T_{m-1} T_{m-1}^\dagger - \vec{u} \vec{u}^\dagger &
-z_m \vec{u} \\
-\bar{z}_m \vec{u}^\dagger & 1 - |z_m|^2 \end{array} \right ].
$$
Working now as in deriving (\ref{pip}) from (\ref{TT1}) shows
\begin{equation}\label{pip2}
 \det ( {\mathbb I}_m  - T_m T_m^\dagger) = (1 - |z_m|^2) \det (  {\mathbb I}_{m-1} - T_{m-1} T_{m-1}^\dagger)
\Big ( 1 - {1 \over 1 - |z_m|^2} \vec{u}^\dagger (  {\mathbb I}_{m-1} - T_{m-1} T_{m-1}^\dagger)^{-1} \vec{u}
\Big ).
\end{equation}
Substituting (\ref{pip2}) in (\ref{pip1}) and proceeding now as in the derivation of (\ref{ICI})
from  (\ref{ICT}) shows
\begin{equation*}
J_{m,p}(z_1,\dots,z_m) = {C}_{m,p} (1 - |z_m|^2 )^{m+p-1}
J_{m-1,p+1}(z_1,\dots,z_{m-1})
\end{equation*}
where $C_{m,p}$ is again given by (\ref{CNP}). Iterating this equation $m$ times, we get
\begin{equation}\label{ICI1}
J_{m,p}(z_1,\dots,z_m) = {C}_{m,p} \prod \limits_{k=1}^{m}(1-|z_{k}|^{2})^{m+p-1}.
\end{equation}

Applying (\ref{ICI1}) with $m=N$, $p=n-N$ shows
$$
J_N(z_1,\dots,z_N) \propto \prod_{l=1}^N (1 - |z_l|^2)^{n-1} 
$$
and this substituted in (\ref{YU2a}) gives (\ref{zx}). \hfill $\square$

\section*{Acknowledgements}
The work of PJF was supported by the Australian Research Council.


\providecommand{\bysame}{\leavevmode\hbox to3em{\hrulefill}\thinspace}
\providecommand{\MR}{\relax\ifhmode\unskip\space\fi MR }
\providecommand{\MRhref}[2]{%
  \href{http://www.ams.org/mathscinet-getitem?mr=#1}{#2}
}
\providecommand{\href}[2]{#2}

\end{document}